# Optimization of a portable liquid scintillation counting device for determining $^{222}$Rn in water

**Santiago Celaya** [1,2], **Ismael Fuente**[1,*], **Luis Quindós**[1,2], **Carlos Sainz**[1,2].

[1] Radon Group, University of Cantabria, C/Cardenal Herrera Oria s/n, 39011, Santander, Spain.

[2] The Cantabrian International Institute for Prehistoric Research (IIIPC), University of Cantabria. Avda de Los Castros nº 52, 39005. Santander, Spain.

**Graphical abstract**

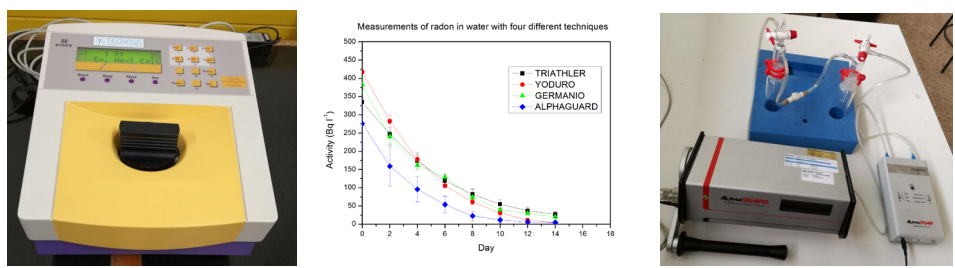

**Abstract** The new EU Council Directive 2013/51/Euratom of 22 October 2013 introduced limits for the content of $^{222}$Rn in drinking water. Radon analysis in water requires a lengthy task of collection, storage, transport and subsequent measurement in a laboratory. A portable liquid scintillation counting device allows rapid sampling with significant savings of time, space, and cost compared with the commonly used techniques of gamma spectrometry or methods based on the desorption of radon dissolved in water. In this study, we describe a calibration procedure for a portable liquid scintillation counting device that allows measurements of $^{222}$Rn in water by the direct method, and we also consider the case of $^{226}$Ra being present in the sample. The results obtained with this portable device are compared with those obtained by standard laboratory techniques (gamma spectrometry with a high-purity Ge detector, gamma spectrometry with a NaI detector, and desorption followed by ionization chamber detection).

## 1. Introduction

Radon is a natural noble gas with three main isotopes: $^{222}$Rn ($t_{1/2}$ = 3.8232 days; hereinafter referred to as *radon*), $^{220}$Rn ($t_{1/2}$ = 55.8 s, called *thoron*), and $^{219}$Rn ($t_{1/2}$ = 3.98 s, called *actinon*) [1]. The contribution of radon exposure to the total estimated average effective dose is around 50% (1.3 mSv per year) [2]. However, because of its low transfer coefficient from water to air [3], little attention has been paid to the radon content in water.

The concentrations of radon in water depend mainly on the radium content of the substrate, the specific surface area of the aquifer, the permeability of the soil, and the water characteristics [4]. Therefore, when groundwater reaches the surface, radon concentrations decrease rapidly as the water moves and is purified, but if water is consumed directly, the health hazard from ingestion of radon and its progeny may be higher. This risk has led to the need for major changes in European and Spanish legislation. Council Directive 2013/51/Euratom of 22 October 2013 [5] (which sets a range of between 100 and 1000 Bq l$^{-1}$ for radon in drinking water) has been introduced into Spanish legislation [6]. This Spanish legislation establishes a limit for radon concentration in drinking water of 500 Bq l$^{-1}$.

The Laboratory of Environmental Radioactivity of the University of Cantabria routinely uses various techniques to measure radon in water: gamma spectrometry (high-purity Ge [HPGe] and NaI detectors) and a desorption technique to measure radon alpha emitters that was applied to determine $^{226}$Ra in bottled water [7] and $^{222}$Rn in water in a region of high radiation [8]. The downside of these techniques is their limited applicability in fieldwork. The development of new portable devices based on liquid scintillation counting (LSC) can solve this issue [9]. LSC is an effective technique to determine radioisotopes that was developed for beta counting in the late 1950s. It was not until the mid-1960s that it was used for the determination of alpha emitters, because of its high counting efficiency (close to 100%) [10]. Portable LSC devices allow large numbers of samples to be processed, with the advantage of in situ analysis [11].

The aim of this study was to characterize and calibrate a portable LSC device to measure $^{222}$Rn in water samples, with and without the presence of $^{226}$Ra, by application of the direct method. We compare our results with those from standard laboratory techniques (gamma spectrometry and desorption followed by ionization chamber detection).

## 2. Materials and methods

### 2.1. Sample collection

Samples were collected at the thermal spa of Las Caldas de Besaya, whose waters have been studied by our group since the 1990s [12, 13], while additional samples were collected from our laboratory, where radon in water was generated with an experimental setup consisting of a methacrylate box of approximately 0.2 m$^3$ filled with water. Inside, pumped air circulates via a silicone tube that is connected to a source of uranium, ensuring that radon is distributed evenly throughout the water, as can be seen in Fig. 1 [14].

For LSC, the water sample was collected in a glass beaker, and a pipette was used to transfer 6 ml to a 20 ml glass vial (to which 14 ml Aqualight liquid scintillation solution for the direct method had been previously added). Care was taken to introduce the sample below the scintillator fluid, with manual agitation of the vial for at least 30 s to allow complete mixing.

For gamma spectrometry with a NaI detector, the water sample was placed in a cylindrical plastic bottle (diameter 6.5 cm and height 10 cm; approximate volume 280 ml), while for gamma spectrometry with an HPGe detector, the cylindrical bottle used had smaller dimensions (diameter 8 cm and height 6.5 cm; approximate volume 270 ml). In both cases, the bottles were filled so as to minimize the space between the surface of the liquid and the cap, so as to avoid the presence of free air inside the sample bottle; the bottle was then closed with a double cap to prevent leakage.

For the desorption technique, the sample was placed in a glass beaker, and the required volume of 100 ml was transferred to a glass Aquakit cylinder. Fig. 2 shows the specific containers used for each technique.

### 2.2 Measurement techniques

#### 2.2.1. LSC with alpha/beta separation

The LSC device used was a small Triathler (model 425-034; 330 × 250 × 190 mm) with an integrated alpha/beta separator (Hidex, Finland). Alpha/beta separation is a feature of the Triathler: it applies a pulse length index that differentiates the longer pulse duration of alpha particles (≃100 ns) from the shorter pulses of beta particles (<30 ns) [15]. This technique detects the alpha emissions of $^{226}$Ra, $^{222}$Rn, and its progeny by means of a photomultiplier tube [16]. The liquid scintillator used (Aqualight, Hidex) is a hydrocarbon with two aromatic rings known as *diisopropyl naphthalene*, which is used in the so-called direct method to measure radon in water.

Eq. (1) was used to calculate the activity due to radon in the water 3 h after the sample and scintillator had been prepared in the vial.

$$A = \frac{G-B}{E_f \cdot 60 \cdot V}, \qquad (1)$$

where $A$ is the activity (Bq l$^{-1}$), $G$ is the counts per minute (cpm), $B$ is the background count obtained with the equipment for a sample prepared with distilled water (2 cpm in this study), $E_f$ is the equipment's efficiency (counts per second/disintegrations per second), $V$ is the sample volume (l), and 60 is a factor to transform counts per minute to counts per second. The elapse time of 3 h is necessary to achieve secular equilibrium between radon and its short-lived progeny ($^{218}$Po, $^{214}$Pb, $^{214}$Bi, and $^{214}$Po).

*LSC device calibration for the direct method*

The initial efficiency (of the first measurement) was 2.80 (as given in the owner's handbook: 0.93 per radioisotope present) when only $^{222}$Rn and its progeny were present in the sample (see Fig. 3). In contrast, an efficiency of 3.73 was used with a sample containing $^{226}$Ra (see Fig. 4).

In Fig. 3, the peak in the window labeled *a* from channels 400-630 corresponds to $^{222}$Rn and $^{218}$Po, while window *b*, from channels 630-800, identifies $^{214}$Po. The peak in Fig. 4 from window *c*, from channels 320-600, is wider than peak *a* because of the presence of $^{226}$Ra, $^{222}$Rn, and $^{218}$Po, while in peak *d* there is only $^{214}$Po.

The first step to set the efficiency started with a certified source of $^{226}$Ra supplied by the National Institute of Standards and Technology (NIST) with an activity of 2482 Bq g$^{-1}$, according to its certificate. From this source, dilutions were made with high-purity water at various known concentrations: 12, 20, 32, 100, 296, 843, and 2943 Bq l$^{-1}$. The vials corresponding to these dilutions are called *G, F, E, D, C, B*, and *A* in the tables.

The second step was to fill a vial for each concentration, by addition of 6 ml of each dilution below the 14 ml of liquid scintillator previously added. These vials were left for 40 days before measurement to allow full ingrowth of the radon progeny [17].

After this time the vials were measured, yielding the results presented in Table 1, where the Triathler value is the result obtained for each vial by application of Eq. (1) with $E_f = 3.73$.

The results obtained from application of the efficiency given in the manual (3.73 when radium is present) show differences from the theoretical values that are impossible to solve by changing only the efficiency. The explanation lies in the interference generated by $^{210}$Po (alpha emitter, 138.3763 days), which is distant descendant of $^{226}$Ra found in standard samples of $^{226}$Ra that have not been purified in the last 10 years. (The certificate for the standard sample used in our study indicates that it has not been purified for 67 years.)

Ten years after purification, the activity of the $^{210}$Pb radioisotope is estimated to be around 27% of that established for $^{226}$Ra [18]. To estimate the error due to the presence of $^{210}$Po (the same activity as for $^{210}$Pb) in the calibrated samples prepared in this study and to determine the efficiency, Eq. (2) was applied to determine the theoretical activity of this radioisotope considering the date of preparation of the NIST certified source:

$$A_{Po} = A_{Bi} = A_{Pb} = A_{Ra}(1 - e^{-\lambda t}), \quad (2)$$

where $A_{Ra}$ is the activity of $^{226}$Ra in the sample (disintegrations per minute, dpm), $A_{Po}$ is the activity of $^{210}$Po in the sample (dpm), $A_{Pb}$ is the activity of $^{210}$Pb in the sample (dpm), $A_{Bi}$ is the activity of $^{210}$Bi in the sample (dpm), $\lambda$ is the decay constant of $^{210}$Pb (0.693/22.23 per year), and $t$ is the time from $^{226}$Ra-standard purification to the calibration (years).

Table 2 shows in the last column the estimated values of $^{210}$Po (the same value as for $^{210}$Pb) obtained with Eq. 2 for each vial, while the middle column presents the theoretical values of $^{226}$Ra in the 6 ml of known samples in each vial.

Once the activity of $^{210}$Po ($A_{Po}$) in each vial had been estimated (last column of Table 2), the total alpha activity ($A_\alpha$) was determined (Table 3) by multiplication of the theoretical activity of $^{226}$Ra by 4 (since $^{226}$Ra, $^{222}$Rn, $^{218}$Po, and $^{214}$Po are in equilibrium) and addition of this to the estimate for $^{210}$Po.

Table 3 presents the values of $A_\alpha$ for each vial against the total alpha count quantified by the Triathler. Fig. 5 shows a graphical representation of the values in Table 3.

The linear fit in Fig. 5 corresponds to Eq. (3), whose slope is the efficiency in this calibration:

$$C_{\text{Total}} = 8.70 + 0.87 \cdot A_\alpha, \qquad (3)$$

Table 4 shows the activities calculated by application of an efficiency of 0.87 per radioisotope present in the spectrum to the counts measured by the Triathler. These values are similar to the theoretical activities of the samples prepared with the NIST certified source, taking into account their uncertainties.

With use of the aforementioned results and a value of 0.87 per radioisotope, a definitive efficiency of 2.62 (three radioisotopes × 0.87) is established when the sample contains radon without the presence of radium (Fig. 3), while it is 3.49 (four radioisotopes × 0.87) when the spectrum (Fig. 4) indicates the presence of $^{226}$Ra.

The good results obtained by our laboratory in national [19] and international intercomparison exercises for radon in water demonstrates that the efficiency calculated was correct.

**2.2.2. Gamma spectrometry**

*2.2.2.1. HPGe coaxial detector*

The equipment used in this study was an HPGe coaxial detector (model GL-2015-7500, Canberra, USA). The device is designed to detect gamma emissions from soil, sludge, ash, environmental filters and, ultimately, any sample whose gamma emission falls between 30 and 3000 keV. The photons resulting from gamma emissions from the sample enter the active volume of the detector and interact with its atoms. These interactions are converted to electrical pulses that are proportional to the energy of the photons emitted, and which are stored in equivalent finite energy increments over the range of the spectrum [20].

$^{222}$Rn activity was determined 3 h after preparation of the sample bottle, with the count performed in the area of the spectrum corresponding to $^{214}$Pb (351.932 keV), as shown in Fig. 6 where window *e* contains the peak of $^{214}$Pb for this emission energy. A 3

h elapse time is necessary to achieve secular equilibrium between radon and its progeny ($^{218}$Po, $^{214}$Pb, $^{214}$Bi, and $^{214}$Po).

*2.2.2.2. NaI detector*

The equipment used was a NaI detector (Canberra, USA), which uses a scintillation technique based on the luminescence radiation produced by interaction with certain materials. Depending on the material used, the scintillation detector can be liquid or solid. In this case a solid scintillation detector was used for spectrometric determination of gamma emitters. $^{222}$Rn activity was measured by the counts measured in window *f*, as seen in Fig. 7, corresponding to the triplet of $^{214}$Pb (241.997, 295.224, and 351.932 keV), and window *g*, corresponding to $^{214}$Bi (609.312 keV) [1].

## 2.2.3 Desorption technique for $^{222}$Rn in water followed by ionization chamber detection

The equipment used was an AlphaGuard PQ2000-PRO (Bertin Instruments, France), which uses a specific kit for measuring $^{222}$Rn in water. The system shown in Fig. 8 allows continuous bubbling (generated by the pump), which causes desorption of radon in the water and directs it to the detector via a security vessel. Once inside the detector, the radon diffuses and passes through a large-surface-area fiberglass filter (which prevents entry of its progeny and aerosols) into a cylindrical ionization chamber (where a potential of 750 V is maintained). The alpha particles emitted by radon ionize the air, causing the cathode to attract positively charged particles and the anode to attract negatively charged particles.

## 3. Results and discussion

### 3.1. Water samples with $^{226}$Ra in the Triathler

For water samples where $^{226}$Ra is present (detected from an alpha spectrum such as in Figs. 3 and 4), it is necessary to wait at least 40 days for the $^{222}$Rn and its progeny to grow and reach secular equilibrium [17]. Fig. 9 shows the alpha spectrum of a water sample containing radium over a period of 40 days. The sample prepared in our laboratory was bubbled with air overnight to eliminate all radon and then measured from the first day (when it contains only $^{226}$Ra) to the 40th day (when it contains $^{226}$Ra, $^{222}$Rn, $^{218}$Po, and $^{214}$Po).

When the sample water contains $^{226}$Ra, there are three options for determining the $^{222}$Rn activity by the direct method with the Triathler:

1. The sample is measured on day 1 after bubbling overnight. This sample contains only $^{226}$Ra (the efficiency per radioisotope is 0.87). The measurement of 437 cpm was put into Eq. (1), which predicts a $^{226}$Ra concentration of 1386 ± 96 Bq l$^{-1}$, which will be the $^{222}$Rn concentration after 40 days, when secular equilibrium is reached.

2. The sample is measured after 40 days, once secular equilibrium has been reached. The measurement in this case was 1755 cpm, corresponding to the four radioisotopes in equilibrium ($^{226}$Ra, $^{222}$Rn, $^{218}$Po, and $^{214}$Po). From Eq. (1), assuming an efficiency of 3.49 (0.87 per radioisotope), a concentration of $^{222}$Rn of 1395 ± 114 Bq l$^{-1}$ is obtained.

3. The sample is measured after 40 days and the counts for $^{226}$Ra are subtracted: 1755 - 437 = 1318 cpm, which corresponds to the triplet $^{222}$Rn, $^{218}$Po, and $^{214}$Po. The result from Eq. (1) with an efficiency of 2.62 was 1396 ± 120 Bq l$^{-1}$.

These results are presented in Table 5, which confirms that the three options give the same results. Option 1 is the fastest; however, it requires a pump to bubble the sample overnight. If a 40-day wait is acceptable, option 2 is the easiest. Option 3 is ideal when the counts of only $^{222}$Rn and its progeny are required.

### 3.2. Performance of the four techniques on water samples without $^{226}$Ra

Table 6 shows the results obtained with the four techniques for samples collected in the thermal spa and those generated in our laboratory setup.

Table 7 shows the analysis of the results by means of a paired *t* test. The results from the Triathler, HPGe, and NaI methods were not significantly different, while the results from the AlphaGuard technique were markedly different.

### 4. Conclusions

The calibration of the portable LSC device, the Triathler 425-034, performed in this study by use of the direct method and water samples containing known concentrations of $^{226}$Ra yielded very good results. This was verified by the results of the intercomparison exercises for $^{222}$Rn in water in which we participated in November 2015 and December 2016 [19].

The current study presents the possibility of determining the $^{222}$Rn activity of a water sample when $^{226}$Ra appears in its alpha spectrum, without having to wait for the secular equilibrium with its progeny to be reached ($\simeq$40 days). The results obtained by LSC with the three options were similar, and the best option depends on the time available to deliver the results.

The Triathler, HPGe, and NaI methods present similar results, and any of these methods are suitable to measure $^{222}$Rn in water. Nevertheless, for rapid measurements that give significant savings of time, space, and cost, the LSC technique is the best because it requires a smaller sample (6 ml), and short measurement interval (600 s), and the measurement can be made in situ at the moment the sample is taken.

The desorption followed by ionization technique (AlphaGuard) yields results different from those from the other three techniques not only in this study but also in the intercomparison exercises undertaken by Iproma S.A and the University of Cantabria involving 17 Spanish laboratories using various techniques [19].

20. Fuente Merino, I. *Puesta a punto de un equipo de fluorescencia de rayos x portátil con fuentes radiactivas: Aplicaciones medioambientales*. Thesis, University of Cantabria (2015). https://repositorio.unican.es/xmlui/bitstream/handle/10902/6787/Tesis%20IFM.pdf.

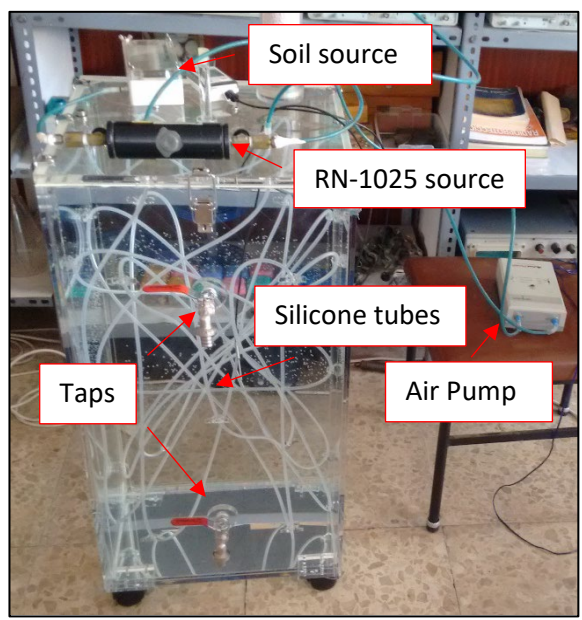

**Fig. 1**. Artificial generation of radon in water.

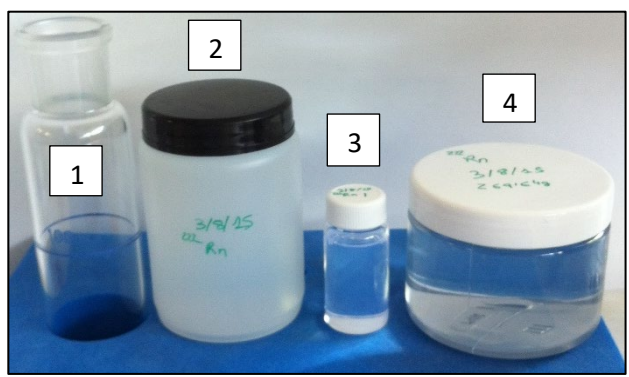

**Fig. 2**. Specific containers used for each technique: 1, desorption; 2, NaI detector; 3, liquid scintillation counting; 4, high-purity Ge detector.

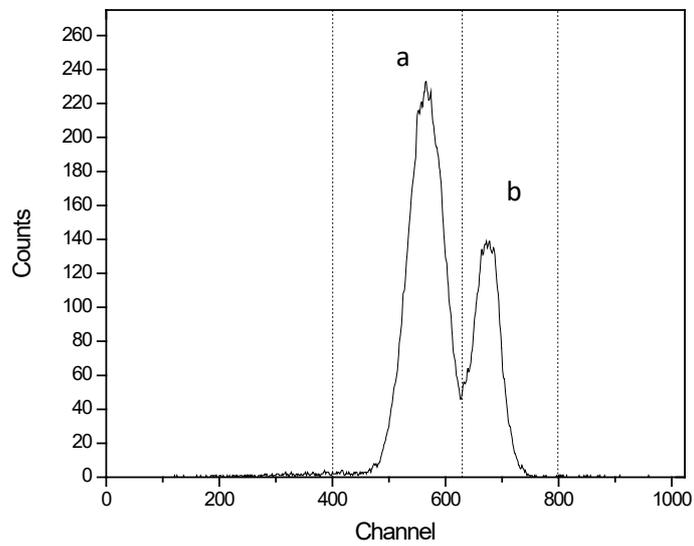

**Fig. 3.** Alpha spectrum of sample with $^{222}$Rn, $^{218}$Po, and $^{214}$Po.

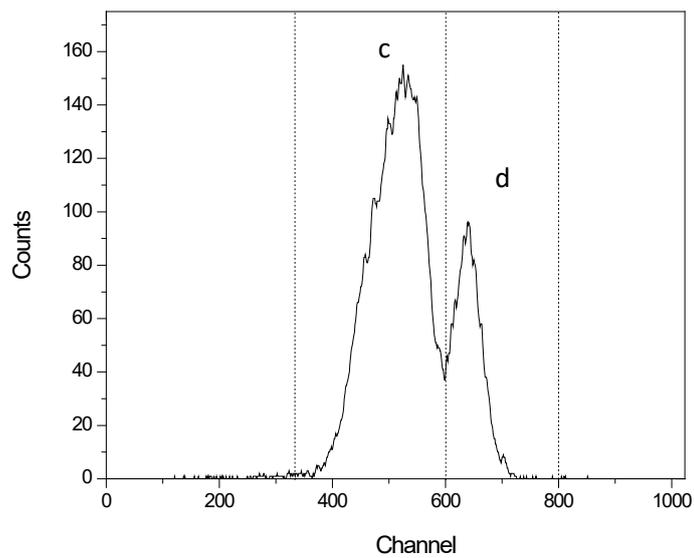

**Fig. 4.** Alpha spectrum of sample with $^{226}$Ra, $^{222}$Rn, $^{218}$Po, and $^{214}$Po.

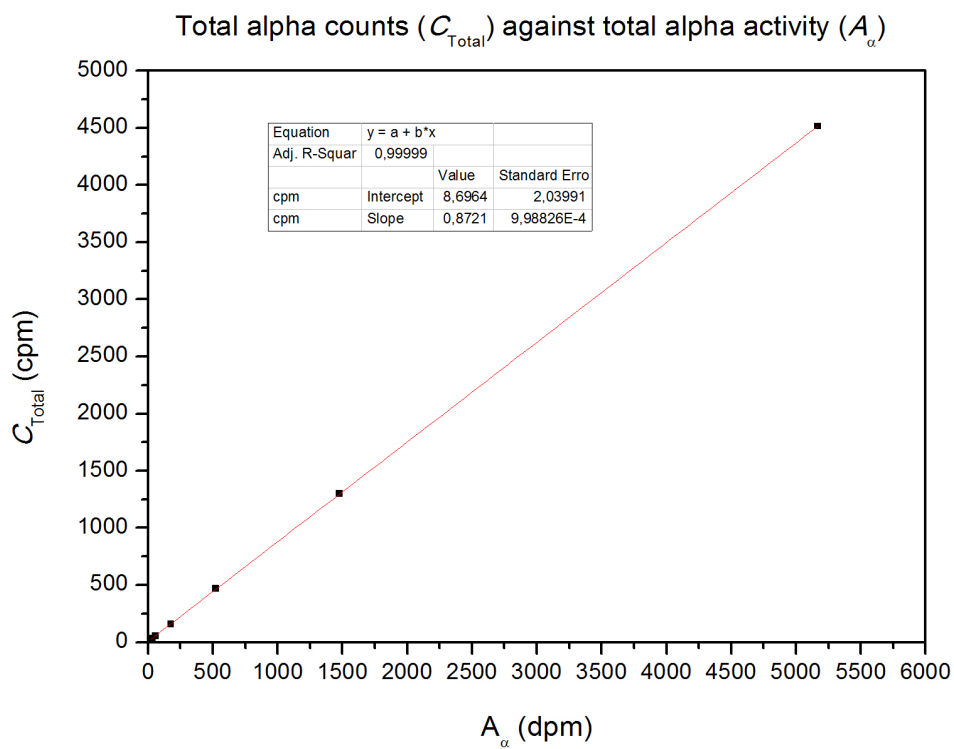

**Fig. 5.** Total alpha counts in counts per minute (cpm) against total alpha activity in disintegrations per minute (dpm).

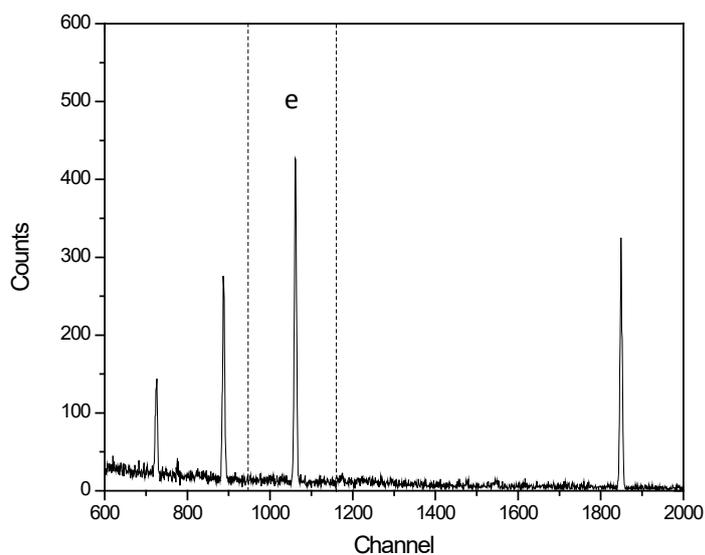

**Fig. 6.** Gamma spectrum obtained with the high-purity Ge detector of a water sample with $^{222}$Rn.

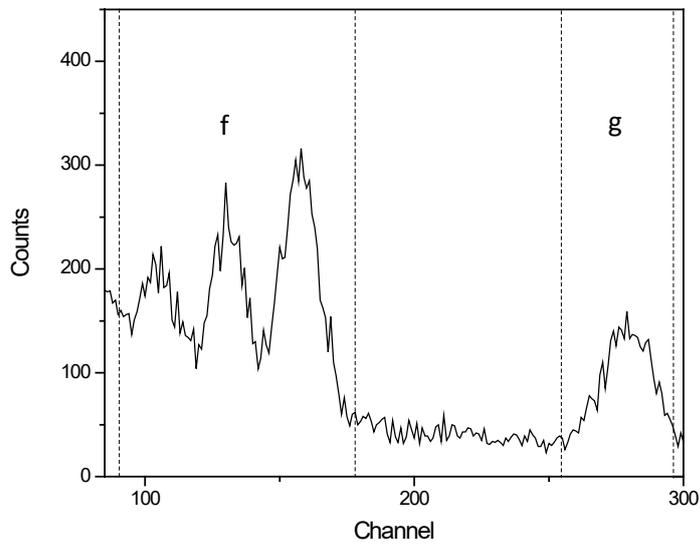

**Fig. 7.** Gamma spectrum obtained with the NaI detector of a water sample with $^{222}$Rn.

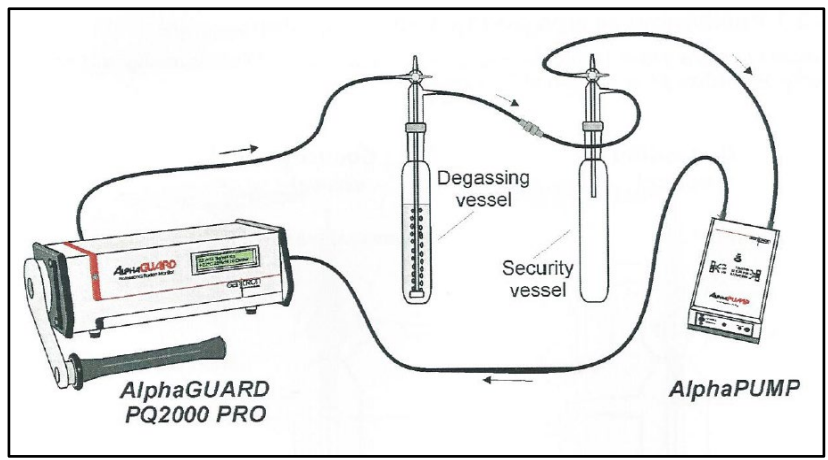

**Fig. 8.** AlphaGuard with a specific $^{222}$Rn kit in water.

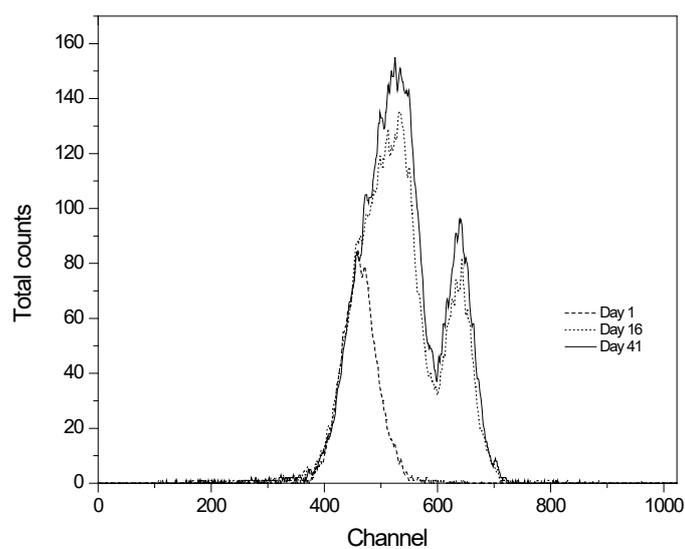

**Fig. 9.** Alpha spectrum over time of a water sample containing only $^{226}$Ra on day 1 (after bubbling), until day 41 (when it contained $^{226}$Rn, $^{222}$Rn, $^{218}$Po, and $^{214}$Po).

**Table 1.** Comparison between the theoretical value and the Triathler value with use of the efficiency (3.73) given in the handbook.

| Vial | Theoretical value (Bq l$^{-1}$) | Triathler value (Bq l$^{-1}$) |
|:---:|:---:|:---:|
| A | 2943 ± 141 | 3360 ± 274 |
| B | 843 ± 41 | 969 ± 88 |
| C | 296 ± 14 | 350 ± 38 |
| D | 100 ± 5 | 120 ± 18 |
| E | 32 ± 2 | 42 ± 10 |
| F | 20 ± 2 | 26 ± 8 |
| G | 12 ± 1 | 18 ± 6 |

**Table 2.** Value estimated for $^{210}$Po in each vial by Eq. (2).

| Vial | Theoretical value for $^{226}$Ra (Bq l$^{-1}$) | Theoretical value for $^{226}$Ra (dpm) | Estimated Value for $^{210}$Po (dpm) |
|---|---|---|---|
| A | 2943 ± 141 | 1060 ± 51 | 928 ± 25 |
| B | 843 ± 41 | 303 ± 15 | 265 ± 13 |
| C | 296 ± 14 | 107 ± 5 | 93 ± 4 |
| D | 100 ± 5 | 36 ± 2 | 32 ± 2 |
| E | 32 ± 2 | 12 ± 1 | 10 ± 1 |
| F | 20 ± 2 | 7 ± 1 | 6 ± 1 |
| G | 12 ± 1 | 4 ± 1 | 4 ± 1 |

dpm, disintegrations per minute

**Table 3.** Total alpha activity ($A_\alpha$) in disintegrations per minute and total alpha counts ($C_{Total}$) in counts per minute for each vial.

| Vial | $A_\alpha = 4 \cdot A_{Ra} + A_{Po}$ | $C_{Total}$ |
|---|---|---|
| A | 5168 ± 2206 | 4514 |
| B | 1477 ± 61 | 1302 |
| C | 521 ± 20 | 470 |
| D | 176 ± 8 | 161 |
| E | 58 ± 4 | 56 |
| F | 34 ± 4 | 35 |
| G | 20 ± 4 | 24 |

**Table 4.** Comparison between the theoretical and estimated values with the calculated efficiency (0.87).

| Vial | Theoretical value (Bq l$^{-1}$) | Estimated value (Bq l$^{-1}$) |
|---|---|---|
| A | 2943 ± 141 | 2944 ± 236 |
| B | 843 ± 41 | 850 ± 72 |
| C | 296 ± 14 | 309 ± 28 |
| D | 100 ± 5 | 106 ± 12 |
| E | 32 ± 2 | 32 ± 2 |
| F | 20 ± 2 | 20 ± 2 |
| G | 12 ± 1 | 12 ± 1 |

**Table 5.** Results obtained with the Triathler for water samples with $^{226}$Ra.

| Option | Measurement date | Radioisotopes | Counts per minute | Efficiency | Activity (Bq l$^{-1}$) |
|---|---|---|---|---|---|
| 1 | Day 1 | $^{226}$Ra | 437 | 0.87 | 1386 ± 96 |
| 2 | Day 40 | $^{226}$Ra, $^{222}$Rn, $^{218}$Po, and $^{214}$Po | 1755 | 3.49 | 1395 ± 114 |
| 3 | Day 40 | $^{222}$Rn, $^{218}$Po, and $^{214}$Po | 1318 | 2.62 | 1396 ± 120 |

**Table 6.** Results (Bq l$^{-1}$) and uncertainties ($k = 2$) obtained for the water samples with each of the four techniques.

| Day | Triathler (MT: 600 s) | HPGe (MT: 3600 s) | NaI (MT: 3600 s) | Alphaguard (MT: 600 s) |
|---|---|---|---|---|
| Samples collected in the thermal spa of Las Caldas de Besaya | | | | |
| 0 | 335 ± 36 | 382 ± 24 | 417 ± 24 | 276 ± 71 |
| 2 | 247 ± 30 | 240 ± 15 | 283 ± 17 | 159 ± 54 |
| 4 | 174 ± 22 | 162 ± 11 | 178 ± 12 | 96 ± 35 |
| 6 | 119 ± 18 | 129 ± 9 | 106 ± 8 | 54 ± 23 |
| 8 | 82 ± 14 | 74 ± 5 | 61 ± 7 | 23 ± 13 |
| 10 | 55 ± 12 | 40 ± 4 | 31 ± 6 | 12 ± 9 |
| 12 | 37 ±10 | 30 ± 3 | 11 ± 6 | 6 ± 6 |
| 14 | 27 ± 8 | 20 ± 2 | 5 ± 6 | 5 ± 6 |
| Samples from radon-water generated in our laboratory | | | | |
| 0 | 150 ± 20 | 156 ± 10 | 186 ± 12 | 114 ± 40 |
| 2 | 101 ± 16 | 95 ± 7 | 120 ± 9 | 62 ± 25 |
| 4 | 67 ± 12 | 57 ± 5 | 70 ± 7 | 37 ± 18 |
| 6 | 48 ± 10 | 27 ± 3 | 41 ± 6 | 21 ± 12 |
| 8 | 33 ± 8 | 18 ± 2 | 29 ± 6 | 12 ± 10 |
| 10 | 24 ± 8 | 10 ± 2 | 17 ± 6 | 7 ± 7 |
| 12 | 14 ± 6 | 8 ± 2 | 5 ± 6 | 4 ± 5 |

HPGe, high-purity Ge; MT, measurement time.

**Table 7**. Results of the paired *t* test for the four techniques.

| | DF | Mean X-Y[a] | Paired *t* value | Probability (two-tail) | Significantly different |
|---|---|---|---|---|---|
| Tr vs Ge | 14 | 4.33 | 1.04 | 0.318 | No |
| Tr vs Na | 14 | -3.20 | -0.42 | 0.679 | No |
| Tr vs Ag | 14 | 41.67 | 6.95 | 6.74E-6 | Yes |
| Ge vs Na | 14 | -7.53 | -1.42 | 0.177 | No |
| Ge vs Ag | 14 | 37.33 | 4.51 | 4.93E-4 | Yes |
| Na vs Ag | 14 | 44.87 | 3.98 | 0.001 | Yes |

[a] Difference between the two means.

Ag, AlphaGuard method; DF, degrees of freedom; Ge, high-purity Ge method; Na, NaI method; Tr, Triathler method.